\documentclass[letterpaper,titlepage,11pt]{article}
\usepackage{hyperref}
\usepackage{amssymb,amsmath,amsfonts,enumerate}
\usepackage{graphicx}

\def\clock{{\count0=\time
           \divide\count0 60
           \ifnum\count0<10 0\fi\the\count0
           \multiply\count0 -60 \advance\count0 \time
           :\ifnum\count0<10 0\fi \the\count0
         }}
\newcommand{\timestamp}{{\small\vbox{\hbox{\tt\jobname.tex}
\hbox{\the\day/\the\month/\the\year, \clock}}}}

\newcommand{\beq}{\begin{equation}}
\newcommand{\eeq}{\end{equation}}
\newcommand{\ben}{\begin{displaymath}}
\newcommand{\een}{\end{displaymath}}
\newcommand{\beqa}{\begin{eqnarray}}
\newcommand{\eeqa}{\end{eqnarray}}
\newcommand{\bea}{\begin{eqnarray}}
\newcommand{\eea}{\end{eqnarray}}
\newcommand{\bean}{\begin{eqnarray*}}
\newcommand{\eean}{\end{eqnarray*}}
\newcommand{\ba}{\begin{array}}
\newcommand{\ea}{\end{array}}
\newcommand{\bi}{\begin{itemize}}
\newcommand{\ei}{\end{itemize}}
\newcommand{\ie}{{\it i.e.,\,}}
\newcommand{\eg}{{\it e.g.,\,}}

\numberwithin{equation}{section}

\begin{document}

\begin{titlepage}
\begin{flushright}
\end{flushright}
\vskip 2.cm
\begin{center}
{\bf\LARGE{A New Class of Accelerating Black Hole Solutions}}
\vskip 1.5cm
{\bf Joan Camps$^{a}$, Roberto Emparan$^{a,b}$
}
\vskip 0.5cm
\medskip
\textit{$^{a}$Departament de F{\'\i}sica Fonamental and}\\
\textit{Institut de
Ci\`encies del Cosmos, Universitat de
Barcelona, }\\
\textit{Mart\'{\i} i Franqu\`es 1, E-08028 Barcelona, Spain}\\
\smallskip
\textit{$^{b}$Instituci\'o Catalana de Recerca i Estudis
Avan\c cats (ICREA)}\\
\textit{Passeig Llu\'{\i}s Companys 23, E-08010 Barcelona, Spain}\\

\vskip .2 in
\texttt{jcamps@ub.edu, emparan@ub.edu}

\end{center}

\vskip 0.3in

\baselineskip 16pt
\date{}

\begin{center} {\bf Abstract} \end{center} 

\vskip 0.2cm 

We construct several new families of vacuum solutions
that describe black holes in uniformly accelerated motion. They
generalize the C-metric to the case where the energy density and tension
of the strings that pull (or push) on the black holes are independent
parameters. These strings create large curvatures near
their axis and when they have infinite length they modify the
asymptotic properties of the spacetime, but we discuss how these
features can be dealt with
physically, in particular in terms of `wiggly cosmic strings'. We
comment on possible extensions, and extract lessons for the problem of
finding higher-dimensional accelerating black hole solutions.

\noindent

\end{titlepage} \vfill\eject

\setcounter{equation}{0}

\pagestyle{empty}
\small
\normalsize
\pagestyle{plain}
\setcounter{page}{1}

\newpage

\section{Introduction}

The C-metric is a solution of the Einstein equations that describes the
spacetime of two black holes uniformly accelerating in opposite
directions \cite{cmetric}. This solution and its variants have been
applied to a number of interesting problems, including gravitational
radiation from accelerated sources \cite{radiation}, instantonic pair
creation of black holes \cite{paircreation}, black holes on branes
\cite{bhsonbranes}, five-dimensional black rings \cite{blackrings}, etc.
It is remarkable that a simple, exact solution is available for the
study of such a variety of problems, and so it seems desirable to
investigate possible extensions of it. 

The generalization that we study in this paper can be easily motivated.
In the original C-metric, conical singularities are unavoidable since
they reflect the need of an external force to accelerate the black
holes. Conical deficit angles correspond to distributional string-like
sources with linear energy density $\varepsilon$ and tension
$T=\varepsilon$. The black holes in the C-metric are then accelerated
under the pull of two such semi-infinite strings. These sources are
physically appealing since they can be made sense of as cosmic strings
(vortices) in the limit in which their thickness is negligible. Note,
however, that for the purpose of pulling on the black holes one merely
needs a tensile string. In particular, it is not necessary that its
energy density equals its tension. Thus it appears that a one-parameter
generalization of the C-metric where the energy density of the string is
independent of its tension should be possible. Our purpose is to
describe a solution where black holes are accelerated by such generic
strings.

String-like sources with $\varepsilon\neq T$ arise as the zero-thickness
limit of a variety of less singular sources, such as `wiggly' cosmic
strings or cylindrical shells. A main difference with the
$\varepsilon=T$ sources is that when $\varepsilon\neq T$ the Newtonian
gravitational potential does not vanish and the gravitational field has
non-trivial local curvature. As a consequence the spacetime is not
asymptotically flat, not even locally. Although this may appear as a
serious drawback, it need not be a problem in physical situations in
which the string is not infinite but forms a (possibly long) loop, whose
radius provides a natural cutoff for the geometry at large transverse
distance from the string (this is often assumed also for \eg global
strings). One might also take the view that, like \eg for the Melvin
universe, these strings define their own asymptotic class. We shall
accept that these solutions admit physical motivation.

Interestingly, we can also have `struts' that stretch between the black
holes and push them apart. These solutions are locally inequivalent to
those with pulling strings and since the struts have finite length, the
asymptotic behavior at spatial infinity is expected to be better. The
struts have negative tension, but in contrast to the C-metric, they do
not violate any energy condition if the energy density on the struts is
large enough. There is also the possibility of solutions with both
strings and struts. Although the physical import of these metrics is
less clear, they provide the largest (five-parameter) family of
solutions in this class and we present them explicitly for completeness.

There are several interesting possible extensions of the solutions
described in this paper. In fact one of the reasons leading us to
their study has been the consideration of higher-dimensional
generalizations of the C-metric. In the context of these elusive
solutions, which have been sought for many of the applications mentioned
in our opening paragraph, the possibility of different kinds of string
sources is potentially even more important than in four dimensions. We
shall address this point towards the end of the paper.

In the next section we analyze the basic string solutions that later
will be used to pull, or push, on the black holes. In
sec.~\ref{sec:accbhs} we present the new metrics for accelerating black
holes, first with pulling strings, then with pushing struts, and we
analyze their main properties. We also present the most general solution
with both strings and struts. Sec.~\ref{sec:outlook} discusses
extensions of these solutions. The appendix contains an
alternative form for the new metrics that resembles more closely the
conventional way of writing the C-metric.

\section{The Levi-Civita string and its sources}

The Levi-Civita spacetime\footnote{The solution
does not contain any length scale so the coordinates may be regarded as
normalized relative to an arbitrary length unit.}
\beq\label{LCstring}
ds^2=-\rho^{2m}dt^2+\rho^{2m(m-1)}(dz^2+d\rho^2)+\rho^{2(1-
m)}\frac{d\phi^2}{C^2}\,,
\eeq
is a long-known cylindrically-symmetric solution to the vacuum Einstein
equations \cite{levicivita}. It is a Weyl metric of general Petrov
type\footnote{Except for $m=0,1/2,\pm 1,2$.}, and contains
two parameters: $m$, which determines the local curvature, and $C$,
which introduces a conical structure along the axis when the
normalization of $\phi$ is fixed by identifying $\phi\sim \phi+2\pi$.
When $m=0,1$ with $C=1$ it reproduces Minkowski and Rindler space,
respectively, but for generic values of $m$ and $C$ the solution is not
asymptotically flat as $\rho\to\infty$ and exhibits a curvature
singularity at $\rho=0$. As a Weyl solution, it corresponds to
an infinite line source of the Newtonian potential with linear density
$m/2G$. We shall
refer to it as the `Levi-Civita string'. 

It has been argued that, in order to admit an interpretation as the
spacetime of a cylindrical source, one must have $m<1/2$, or possibly
$m<1$ (see \eg \cite{Belinski:2001ph,bonnor} and references therein).
The precise range
will not concern us much, since later we shall be mostly interested in
small values of $m$, as well as $C$ close to 1, for which the line
source is readily interpreted. Nevertheless let us briefly discuss the
general case. The singular behavior near the axis $\rho=0$ may be
smoothed by replacing the region around it with an extended source, and
a simple example is a cylindrical
tubular shell
\cite{Stachel:1983su,Philbin:1995iz,Wang:1996xd,Herrera:2001sa,Bicak:2002vj,Bicak:2004fw}. 
Cutting the metric at $\rho=\rho_s$ and replacing
the interior with flat Minkowski spacetime ($m=0$, $C=1$), one can apply
Israel's analysis \cite{Israel:1966rt} to obtain the stress tensor at
the shell interface
\beqa\label{shell}
T^t_t&=&\frac{\rho_s^{m-1}}{8\pi G}\left((1-m)^2 
\rho_s^{-m^2}-C\right)\,,\nonumber\\
T^z_z&=&\frac{\rho_s^{m-1}}{8\pi G}\left(
\rho_s^{-m^2}-C\right)\,,\\
T^\phi_\phi&=&\frac{\rho_s^{m-1}}{8\pi G}m^2 \rho_s^{-m^2}\,.\nonumber
\eeqa
In this manner, the problem of interpreting the strong curvature
singularity at $\rho=0$ is shifted to that of finding an adequate source
that smoothens the milder singularity at the shell. 
Observe in
\eqref{shell} the presence of not only energy density and tension along
$z$, but also a hoop stress $T^\phi_\phi$, which seems to be a necessary
feature of any possible source of these spacetimes. Typically the equation of
state of the shell matter will impose a relationship between $m$, $C$, and
$\rho_s$. We shall not dwell
much on candidate shell sources, but merely note that tubular
structures with similar properties appear naturally in string theory in
the form of supertubes and closely related helical strings (smeared
along the $z$ direction). It seems likely that combinations or variants
of these can provide adequate sources for these spacetimes.

From \eqref{shell} we can introduce the energy density per unit length
\cite{Bicak:2002vj},
\beq\label{veps}
\varepsilon=-\int_{\rho=\rho_s} d\phi \sqrt{g_{\phi\phi}}\, T^t_t
=\frac{1}{4G}\left(1-\frac{(1-m)^2}{C\rho_s^{m^2}}\right)
\eeq
and tension
\beq\label{tens}
T=-\int_{\rho=\rho_s} d\phi \sqrt{g_{\phi\phi}}\, T^z_z
=\frac{1}{4G}\left(1-\frac{1}{C\rho_s^{m^2}}\right)\,.
\eeq
These are non-trivially equal only in the case of a conical defect
spacetime, $m=0$, $C\neq 1$.

Let us now consider the Levi-Civita spacetime \eqref{LCstring} expanded
to linear order in $m$ and in $\gamma\equiv
C-1$,
\beqa\label{linmet}
ds^2&\simeq&
-\left(1+2m\log\rho\right)dt^2+\left(1-2m\log\rho\right)(dz^2+d\rho^2)\nonumber\\
&&+
\left(1-2m\log\rho\right) \left(1-2\gamma\right)\rho^2d\phi^2\,.
\eeqa
In this linearized approximation we can also write $m$ and $\gamma$ in
terms of the
energy density \eqref{veps} and tension \eqref{tens} as
\beq\label{mandT}
m\simeq 2G(\varepsilon-T)\,,\qquad \gamma\simeq 4GT\,.
\eeq
We will regard these simple relations as the basic interpretation of the
parameters of the Levi-Civita string.
Observe that $\rho_s$ does not appear in them. Relatedly,
the hoop stress $T^\phi_\phi$ is $O(m^2)$ and therefore it does not
appear in the
linearized approximation.

Using \eqref{mandT}, and performing the coordinate change 
\beq
(1-4G(\varepsilon+T)\log r)r^2=(1-8GT)(1-4G(\varepsilon-T)\log\rho)\rho^2
\eeq
(to the required
expansion order) the metric \eqref{linmet} is brought to the form
\beqa\label{linsol}
ds^2&\simeq& -\left(1+4G(\varepsilon-T)\log r\right)dt^2+
\left(1-4G(\varepsilon-T)\log r\right)dz^2\nonumber\\
&&
+\left(1-4G(\varepsilon+T)\log r\right)(dr^2+r^2 d\phi^2)\,.
\eeqa
This can be recognized as the solution to the
linearized Einstein equations, in transverse gauge,
\beq
\Box\left(h_{\mu\nu}-\frac{h}{2}\eta_{\mu\nu}\right)=-16\pi G T_{\mu\nu}
\eeq
with distributional stress tensor
\beq
T_{\mu\nu}=\mathrm{diag}(\varepsilon,-T,0,0)\delta(x)\delta(y)
\eeq
which confirms our interpretation of $\varepsilon$ and $T$.

The solution \eqref{linsol} has been previously studied in the context
of a different kind of string source, namely `wiggly strings' (see \eg
\cite{vilenkinshellard}). Cosmic strings have Lorentz-invariant worldsheets so
$\varepsilon=T$, but if they acquire a short-distance structure
(wiggles), then when this is averaged it produces an effective linear source
with $\varepsilon\neq T$. This is an appealing physical realization of
this linearized spacetime.

Finally, observe that the tension of the string may be negative, and
hence the string exerts pressure, while satisfying the usual energy
conditions if $\varepsilon$ is large enough. We will refer to this as
the `Levi-Civita strut'. Let us discuss it in the case of small $m$ and
small $C-1$. Eq.~\eqref{mandT} implies that, when $C<1$ and hence
$\gamma<0$,
\beq
\varepsilon\simeq \frac{m-|\gamma|/2}{2G}\,,\qquad
\varepsilon -T\simeq \frac{m}{2G}\,,\qquad
\varepsilon -|T|\simeq \frac{m-|\gamma|}{2G}
\,.
\eeq
Therefore the weak, strong, and dominant energy conditions are all
satisfied if $m>|\gamma|$. While it does not seem possible to realize
these struts in terms of wiggly cosmic strings, one may still obtain
them from tubular shells. They might be elastically unstable due to the
negative tension, but presumably this depends on the specific shell
matter (see \cite{Bicak:2002vj} for energy conditions
on generic shells).

\section{Accelerating Black Holes}
\label{sec:accbhs}

We describe different families of solutions where the black holes
are accelerated either by strings that pull or by struts that push on
them. They all contain the C-metric as a limit. 

\subsection{Pulling with strings}\label{strings}

\begin{figure}
\begin{center}
\includegraphics[width=0.7\textwidth]{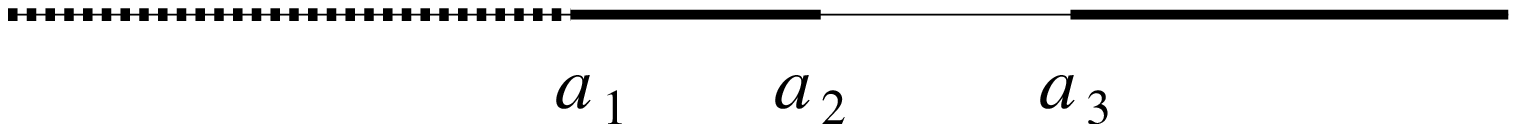}
\caption{Weyl rod structure for the solution with an accelerating black
hole pulled by a semi-infinite Levi-Civita string.
The rod at $z<a_1$ has linear density $m/2G$. The rods at $a_1<z<a_2$
and $z>a_3$ have linear density $1/2G$.
}\label{fig:rods}
\end{center}
\end{figure}

We construct the metric using conventional integrability techniques for
Weyl spacetimes (see \eg \cite{Belinski:2001ph,Emparan:2001wk}). 
The rod structure for the solution\footnote{\textit{I.e.,} the line sources
for the Newtonian potential $U$.} is depicted in fig.~\ref{fig:rods}.
The metric reads
\begin{equation}
ds^2=- e^{2U} dt^2+e^{2\nu}(dz^2+d\rho^2)+e^{-2U}\rho^2 \frac{d\phi^2}{C^2}
\end{equation}
with
\begin{equation}
e^{2U}=\rho^{2m}\,\frac{\mu_1^{1-m}\mu_3}{\mu_2}\,,
\end{equation}
and
\begin{equation}
e^{2\nu}=\rho^{2m(m-1)}\left(\frac{\mu_1}
{(\mu_1^2+\rho^2)^{1-m}}
\left(\frac{\mu_1\mu_2+\rho^2}{\mu_1\mu_3+\rho^2}\right)^2
\right)^{1-m}
\frac{\mu_3(\mu_2\mu_3+\rho^2)^2}
{\mu_2(\mu_2^2+\rho^2)(\mu_3^2+\rho^2)}\,,
\end{equation}
where
\begin{equation}\label{mui}
\mu_i=a_i-z+\sqrt{(a_i-z)^2+\rho^2}\,.
\end{equation}

The solution contains five parameters: $m$, $C$, $a_i$ ($i=1,2,3$). One of the $a_i$
may be absorbed by a shift in $z$ so only their differences $a_i-a_j$ are
physical. The parameter $C$ will be fixed presently by a regularity
condition, so in the end we will be left with a three-parameter family
of solutions. It contains the C-metric as the particular case $m=0$ (see
the appendix), while for $m= +1$ we obtain a double Wick rotation of the
Schwarzschild spacetime. When all the $a_i\to +\infty$ the solution
reduces, after a rescaling of coordinates, to the Levi-Civita spacetime.
Unlike the C-metric, for generic values of the parameters the solution
is not algebraically special.

The rod structure allows a ready interpretation of the
solution. Along the axis
$\rho=0$, we expect to have: 
\begin{itemize}
\item A semi-infinite Levi-Civita string at $z<a_1$ with rod
density $m/2G$. 

\item A black hole horizon at $a_1<z<a_2$. 

\item An `exposed' axis of rotation at $a_2<z<a_3$.

\item An acceleration (Rindler) horizon at $a_3<z$.

\end{itemize}

We proceed to analyze the solution near each of these rods.

\textit{Exposed axis.}
Assuming that $\phi\sim \phi+2\pi$, the absence of conical singularities at
the exposed axis at
$\{\rho=0\,,\,a_2< z< a_3\}$ requires that
\beq
C=\lim_{\rho\to 0}\Bigl.e^{-(U+\nu)}\Bigr|_{a_2< z< a_3}=
\frac{1}{2^m}\frac{(a_3-a_1)^{1-m}}{a_3-a_2}\,.
\eeq

\textit{Black hole horizon.}
Near the rod at $a_1<z<a_2$, to leading order in $\rho$ the metric is
\beqa
ds^2&\simeq&\left(2(z-a_1)\right)^{m-1}\frac{a_3-z}{a_2-z}\left(-\rho^2dt^2+
\left(\frac{a_2-a_1}{a_3-a_1}\right)^{2(1-m)} 
(dz^2+d\rho^2)\right)\nonumber\\
&&+
\frac{a_2-z}{a_3-z}\frac{d\phi^2/C^2}
{\left(2(z-a_1)\right)^{m-1}}\,.
\eeqa
The horizon at $\rho=0$ is a regular surface away from the
Levi-Civita singularity at its pole $z=a_1$.
The horizon area is
\beqa
A_{BH}&=&\int_0^{2\pi}d\phi\int_{a_1}^{a_2}dz\;
\left.\sqrt{g_{zz}g_{\phi\phi}}\right|_{\rho=0}
=\frac{2\pi}{C}(a_2-a_1)\left(\frac{a_2-a_1}{a_3-a_1}\right)^{1-m}
\nonumber\\
&=&
2^{1+m}
\pi(a_3-a_2)\frac{(a_2-a_1)^{2-m}}{(a_3-a_1)^{2-2m}}\,.
\eeqa
We can compute the surface
gravity at the horizon of the Killing vector
$\partial_t$, 
\begin{equation}\label{kbh}
\kappa_{BH}=\lim_{\rho\to
0}
\biggl.\frac{\partial_\rho\sqrt{-g_{tt}}}{\sqrt{g_{\rho\rho}}}
\biggr|_{a_1< z< a_2}=
\left(\frac{a_3-a_1}{a_2-a_1}\right)^{1-m}\,.
\end{equation}
One should keep in mind that given the unusual asymptotics created by the
Levi-Civita string, the normalization of the Killing
generator of the
horizon is somewhat arbitrary.
Observe that the product
\beq\label{smarr}
\kappa_{BH}A_{BH}=\frac{2\pi}{C}(a_2-a_1)
\eeq
is equal to the length of the black hole rod times a factor that
accounts for the modified length of the orbits of $\phi$. One might want
to regard this as a Smarr-type relation $\kappa_{BH}A_{BH}=4\pi GM$ that
would define the mass of the black hole. Actually, this definition of
mass is equivalent to the Komar mass of the black hole on its horizon.
However, it is unclear to what extent this definition of black hole mass
is appropriate in the present context\footnote{See
\cite{Dutta:2005iy}
for a definition of `boost mass'.}.

\textit{Acceleration horizon.}
Near $\rho=0$ and $a_3<z$ we find 
\beq
ds^2\simeq \left(2(z-a_1)\right)^{m-1}\frac{z-a_2}{z-a_3}\left(-
\rho^2 dt^2+dz^2+d\rho^2\right)+
\frac{z-a_3}{z-a_2}\frac{d\phi^2/C^2}
{\left(2(z-a_1)\right)^{m-1}}\,.
\eeq
There is an infinite Killing horizon (Rindler) at $\rho=0$ generated by
$\partial_t$. The apparent singularity at
$z=a_3$ is just a coordinate artifact and the horizon is regular everywhere.
We compute the surface gravity as
was done for the black hole horizon,
\begin{equation}\label{kr}
\kappa_R=1\,.
\end{equation}
The acceleration of the black hole is ambiguous in that it depends on
the normalization of $\partial_t$, which for a spacetime with a
Levi-Civita string is unclear, and also because the black hole is an
extended object. In the case of the C-metric, when the black hole is
small ($a_2-a_1\ll a_3-a_2$) its acceleration relative to static
asymptotic observers can be unambiguously identified to leading order as
$A\simeq (2a_3-(a_1+a_2))^{-1}$.

The ambiguities in the normalization of $\kappa$ cancel when we consider
the quotient
\begin{equation}
\frac{\kappa_R}{\kappa_{BH}}=\left(\frac{a_2-a_1}{a_3-a_1}\right)^{1-m}\,.
\end{equation}
The surface gravities can be associated as usual to horizon
temperatures, $T_{BH,R}=\kappa_{BH,R}/2\pi$.
Since $T_{R}<T_{BH}$ the two temperatures are never equal. Thus, even
if the Levi-Civita singularities at the string and infinity could be
disposed of, an otherwise regular Euclidean instanton could not
be constructed.

\textit{Levi-Civita string.}
For small $\rho$ and $z<a_1$ we have
\beqa
e^{2U}&\simeq& \rho^{2m}\: 2^{1-m}\frac{(a_1-z)^{1-m}(a_3-z)}{a_2-z}\,,
\nonumber\\
e^{2\nu}&\simeq& \rho^{2m(m-1)}\:\frac{a_3-z}{a_2-z}
\left(2^{2m-1}\frac{(a_1-z)^{2m-1}(a_2-z)^2}{(a_3-z)^2}\right)^{1-m}\,.
\eeqa
The radial dependence is like in \eqref{LCstring}, but there
is a dependence on $z$ as well. However, away
from the string endpoint at $z=a_1$ these functions vary slowly with
$z$. Thus let us introduce, at any given
$z$ along the string, the functions $\hat U(z)$ and $\hat\nu(z)$ by
\beqa
e^{2\hat U(z)}&=&2^{1-m}\frac{(a_1-z)^{1-m}(a_3-z)}{a_2-z}\,,\nonumber\\
e^{2\hat\nu(z)}&=&\frac{a_3-z}{a_2-z}
\left(2^{2m-1}\frac{(a_1-z)^{2m-1}(a_2-z)^2}{(a_3-z)^2}\right)^{1-m}\,.
\eeqa
These are approximately constant in a neighbourhood of a given $z$ not
close to $a_1$, so we may locally absorb them through a change of
coordinates in such a way that the geometry is well approximated by a
metric of the form \eqref{LCstring} with a
$z$-dependent $C$ parameter\footnote{We may equivalently say
that we are matching the metrics induced on a surface at constant
$z<a_1$.}
\beq
\hat C(z)=C e^{\hat U(z)} e^{\hat\nu(z)\frac{1-m}{m^2-m+1}}\,.
\eeq
This is a monotonically decreasing function of $z$. In this sense, we
may say that the string tension increases along the string from infinity
towards the black hole.
Note that for the C-metric with $m=0$ this $z$-dependence
cancels out. For small $m$ 
\beq
\hat C(z)=\frac{a_3-a_1}{a_3-a_2}\left(1-m\log\frac{(a_2-z)(a_3-a_1)}{(a_1-
z)(a_3-z)}+O(m^2)\right)\,.
\eeq
It is tempting to interpret this effect as saying that the wiggles in
the string get stretched when this pulls on the black hole, but we have not
pursued this interpretation further.

While it does not seem feasible to identify in a unique manner the mass
and acceleration of the black hole, we note that when the black hole-rod
length $a_2-a_1$ is small (much smaller than $a_3-a_2$) and $m$ is
small, on general grounds we might expect to identify the ratio of
surface gravities as
\beq
\frac{\kappa_R}{\kappa_{BH}}\simeq 4GMA\,.
\eeq
In this limit $C$ is close to 1 so we can identify the string tension
from \eqref{mandT}. If $\varepsilon-T\ll T$ we find that Newton's second law
\beq
T\approx MA
\eeq
is recovered. It is interesting to observe that for the C-metric ($m=0$)
the identity $1-C^{-1}=\kappa_R/\kappa_{BH}$ is exactly satisfied.

\subsection{Pushing with struts}\label{struts}

\begin{figure}
\begin{center}
\includegraphics[width=0.7\textwidth]{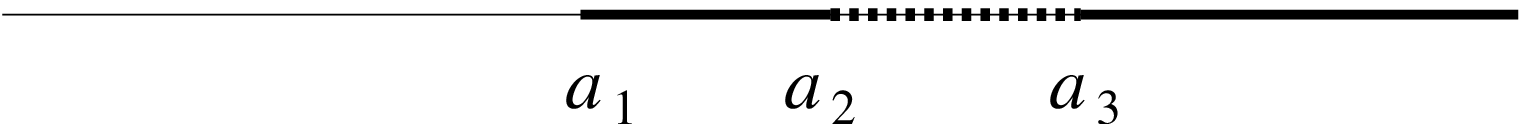}
\caption{Weyl rod structure for the solution with a finite Levi-Civita strut
pushing the black hole.
The rod at $a_2<z<a_3$ has linear density $m/2G$. The rods at $a_1<z<a_2$
and $z>a_3$ have linear density $1/2G$.
}\label{fig:rods2}
\end{center}
\end{figure}
Now we place a finite Levi-Civita rod along $a_2<z<a_3$ while leaving the
semi-infinite axis $z<a_1$ exposed. If the latter is non-singular, then,
as we will see below,
the segment $a_2<z<a_3$ must support a conical excess angle instead of a
deficit angle, and hence we find a Levi-Civita strut pushing on the
black holes. This configuration has the advantage that, since the strut
has finite length, we expect the metric to be asymptotically flat at
spatial infinity. Furthermore, as discussed at the end of the previous
section, the positive pressure along the strut need not imply a
violation of energy conditions as long as $m$ is sufficiently large.
This is unlike in the C-metric with a strut, which always violates
positivity of energy and therefore makes the solutions
manifestly unphysical.

The rod structure is as in fig.~\ref{fig:rods2}. 
The metric functions are
\begin{equation}
e^{2U}=\mu_1\left(\frac{\mu_3}{\mu_2}\right)^{1-m}\,,
\end{equation}
and
\begin{equation}
e^{2\nu}=\left(\frac{\mu_3}{\mu_2}
\left(\frac{\mu_1\mu_2+\rho^2}{\mu_1\mu_3+\rho^2}\right)^2
\left(\frac{(\mu_2\mu_3+\rho^2)^2}{(\mu_2^2+\rho^2)(\mu_3^2+\rho^2)}\right)^{1-m}
\right)^{1-m}
\frac{\mu_1}{\mu_1^2+\rho^2}\,.
\end{equation}
In contrast to the C-metric, when $m\neq 0$ the geometry is locally
inequivalent to our previous solution where the strings run to infinity.
Along $\rho=0$ we now have an exposed axis at $z<a_1$, a black hole
horizon at $a_1<z<a_2$, a Levi-Civita strut with rod density $m/2G$ at
$a_2<z<a_3$ and an acceleration horizon at $a_3<z$.

The area and surface gravities of the horizons take the same form as
in eqs.~\eqref{kbh}, \eqref{smarr}, \eqref{kr}, but now 
regularity at the exposed axis $z<a_1$ requires
\begin{equation}
C=1\,.
\end{equation}
We can then expect an excess angle along the
Levi-Civita rod.
Indeed, for small $\rho$ and $a_2<z<a_3$ we find
\beqa
e^{2U}&\simeq& \rho^{2m}\; 2^{1-2m}\frac{\left((a_3-z)(z-a_2)\right)^{1-m}}{z-a_1}\,,
\nonumber\\
e^{2\nu}&\simeq& 
\rho^{2m(m-1)}\frac{2^{-1+2m-2m^2}}{(z-a_1)^{2m-1}}
\left(
\frac{(a_3-a_2)^{2-2m}((a_3-z)(z-a_2))^{-1+2m}}
{(a_1-a_3)^2}
\right)^{1-m}
\eeqa
and we can define a $z$-dependent $C$ parameter along this rod like
we have done above. For small $m$
we find
\begin{equation}
\hat{C}(z)=\frac{a_3-a_2}{a_3-a_1}\left(1+
m\log\frac{(a_3-a_1)(a_3-z)(z-a_2)}{(z-a_1)(a_3-a_2)^2}+O(m^2)\right)\,,
\end{equation}
which is smaller than $1$, reflecting the need of a negative tension
(pressure) to push the black holes. The rest of the analysis can be
carried out as in the previous solution and we omit it. 

\subsection{Strings and struts}

Clearly, one can construct a larger class of metrics with a Levi-Civita
rod at $z<a_1$ with density $m_L/2G$ and another rod at
$a_2<z<a_3$ with density $m_R/2G$. The construction of these solutions is
straightforward, and the metric functions are
\begin{equation}
e^{2U}=\rho^{2 m_L}\mu_1^{1-m_L}\left(\frac{\mu_3}{\mu_2}\right)^{1-m_R}
\end{equation}
and
\begin{equation}
e^{2\nu}=\left(\frac{\mu_3}{\mu_2}\right)^{1-m_R} 
\left(\frac{(\mu_2\mu_3+\rho^2)^2}{(\mu_2^2+\rho^2)(\mu_3^2+\rho^2)}\right)^{(1-m_R)^2} 
\left(\frac{\mu_1 \rho^{-2m_L}}{(\mu_1^2+\rho^2)^{1-m_L}}  
\left(\frac{\mu_1\mu_2+\rho^2}{\mu_1\mu_3+\rho^2}\right)^{2(1-m_R)}\right)^{1-m_L}\,.
\end{equation}
Since there is no exposed axis there does not seem to be any preferred
value for the parameter $C$. This is then a five-parameter family of
solutions. Their analysis does not introduce any other important
novelties so we shall not dwell on it.

\section{Outlook}
\label{sec:outlook}

We have exhibited several new families of explicit solutions that
describe black holes accelerating under the pull or push of a
string-like object. Their construction is fairly straightforward and our
aim has been to underscore that these solutions can have physical
significance, in particular when strings pull on the black holes. One
important feature is that, even if the Levi-Civita string (or strut) is
strongly singular, it can end on the black hole without destroying the
regularity of the horizon (away from the touchpoint). This feature was
not {\it a priori} obvious, but it follows essentially from the properties of
Weyl rod structures: close to a `horizon rod' the geometry is always of
Rindler type (as we have explicitly exhibited). This is indeed the
reason that, while we have not performed a detailed analysis of the
extension of the solutions across the horizons, we do expect that this
poses no difficulty. 
When the self-gravity of the string is weak (and hence the
acceleration is small) it can be regarded as the zero-thickness limit of
a wiggly cosmic string, but it may also correspond to other non-singular
sources. The main difficulties in interpreting this solution and
identifying its physical parameters stem from its unconventional
asymptotics. But this is a problem only if we consider the string to be
infinitely long, and if the solution is taken to approximate only a
portion of a closed loop of string then the asymptotic behavior will be
improved. On the other hand, the solutions with finite struts
are presumably spatially asymptotically flat.

The existence and properties of these solutions raise a number of
suggestions for future work: 

\paragraph{String sources and cylindrical shells in the accelerating
black hole solution.} We have not investigated the
regularization of the Levi-Civita string in the accelerating black hole
spacetime, but there are reasons to expect that this should not be
problematic. In terms of wiggly
strings, looking sufficiently close to the black hole one may resolve
the wiggles and use the analysis of \cite{Achucarro:1995nu} to conclude
that the vortex string can pierce the black hole. It would remain to
solve the problem of how the wiggly structure extends to all the length
of the string, possibly with $z$-dependent effective parameters as
suggested by our analysis above. One may also replace the Levi-Civita
string with a tubular shell. It does seem possible to cut the solution
at some $\rho=\rho(z)$ in a region $z\leq z_s$, with $z_s<a_2$, and replace
the interior with a smooth spacetime, so the Levi-Civita string is
replaced by an empty cylindrical shell that ends on the black hole.
Israel's construction will yield the shell stress tensor. Stationarity
demands that it be orthogonal to the null generator of the horizon $k$,
\ie\ $k^\mu k^\nu T_{\mu\nu}=0$. Other than this, in the absence of a
specific model for the shell there do not seem to be any
restrictions on its stress tensor.

\paragraph{Black hole charge and pair creation.} The impossibility of
matching the black hole and acceleration temperatures prevents the
construction of a Euclidean instanton that would mediate the snapping of
the string by spontaneous formation of a pair of black holes at its
endpoints. Extending our solution to include black hole charge should
allow one to lower the black hole temperature to match the acceleration
temperature, as in \cite{paircreation}, and then study this process.
For black holes in Kaluza-Klein theory the construction of this
solution should be rather straightforward given the integrability
of the five-dimensional equations.

\paragraph{AdS and black holes on branes.} There does not seem to exist
any obstacle of principle to extending our solutions to include a
(negative) cosmological constant, even if in practice finding exact
solutions might not be feasible (for instance, inverse scattering
techniques are unavailable for this case). At any rate, with these
solutions one could investigate extensions of the construction of
\cite{bhsonbranes} of black holes localized on a Randall-Sundrum
two-brane. Note, however, that the existence of solutions to Israel's
junction conditions for a vacuum brane, \ie one with extrinsic curvature
proportional to its induced metric, is not guaranteed. Also, if the
additional parameter in the solutions allowed one to construct a continuous
family of black holes localized on a two-brane, this might seem to
entail a continuous violation of uniqueness of black holes on the brane.
However, although we are not aware of any theorems against this, it is
unlikely to be realized in this manner since the Levi-Civita string
(`hidden behind the brane') presumably makes it impossible to have flat
asymptotics along the brane directions.

\paragraph{Global structure and gravitational radiation.} In this paper
we have not attempted to study the maximal analytic extension and global
structure of these solutions, but there may be more to this than a point
of mathematical rigour. In particular it should be interesting to study
the extension beyond the Rindler horizon of the solution with pulling
strings to describe the `roof' in the Penrose diagram, where the
radiative properties of the spacetime become apparent (see 
the second reference in \cite{radiation}).
The Levi-Civita string is absent from this region and so it may be
interesting to study whether the asymptotic geometry at null infinity is
better behaved. If radiation at infinity can be suitably characterized,
this may provide an interesting extension of the class of boost-rotation
symmetric radiative spacetimes. On the other hand, the solutions with
struts probably have worse asymptotic behavior in the `roof'.

\paragraph{Non-uniform rod density.} The only parameter that must be
fixed in order to avoid singularities on the exposed axis is $C$, which
amounts to a simple rescaling of $\phi$, and which in the linearized
limit corresponds to the string tension. Thus it would seem possible to
construct Weyl solutions analogous to the ones we have studied, with
naked singularities only at the pulling string, where the rod at $z<a_1$
would have $z$-dependent density $m(z)$ (varying in the range $(0,1/2)$
or possibly $(0,1)$), while $C$ remains constant. Obviously the same
could be done with finite struts. In general, explicit
solutions could be found presumably only up to quadratures, but their
properties might perhaps still be analyzable. This would give a
one-function family of accelerating black holes. It is conceivable that
if $m$ approaches zero sufficiently fast as $z\to-\infty$ the asymptotic
behavior might be as in the C-metric.

\paragraph{Accelerating black holes in higher dimensions.}

No exact solution for accelerating black holes in $D>4$ is known.
Ref.~\cite{Kodama:2008wf} solved the perturbation equations for a $D>4$
Schwarzschild black hole to give it uniform acceleration and
found a solution with a distributional linear source
accelerating the black hole. 

Let us reexamine this problem in light of what we have learned in
four dimensions. In the class of metrics we have analyzed, the C-metric
is singled out as the one where the string has a milder singularity at
the axis and also has better-behaved (locally flat) asymptotics. In contrast,
in $D\geq 5$ we would expect the asymptotic behavior to be good for all
string sources with finite energy density and tension: the gravitational
field in directions transverse to the string falls off like
$\sim 1/r^{D-4}$. 
Near the string, however, there are significant
differences between sources.
Requiring the symmetry ${\mathbb R}_t\times {\mathbb R}_z\times SO(D-2)$,
the static, cylindrically symmetric string-like solutions to the vacuum
Einstein equations can be obtained as a particular case of solutions
in \cite{Gibbons:1987ps} (or in $D=5$ by uplifting the
solutions in \cite{Agnese:1985xj}) to find
\beq
ds^2=-f^{\frac{(D-3)\varepsilon-T}{\mu}}dt^2+f^{\frac{(D-3)T-\varepsilon}{\mu}}dz^2
+f^{-\frac{D-5}{D-4}-\frac{\varepsilon+T}{\mu}}dr^2+f^{\frac{1}{D-4}-\frac{\varepsilon+T}{\mu}}
r^2 d\Omega_{(D-3)}
\eeq
with
\beqa
f&=&1-\frac{16\pi G}{(D-4)(D-2)\Omega_{(D-3)}}\frac{\mu}{r^{D-4}}\,,\nonumber\\
\mu&=&\sqrt{(D-4)(D-2)\left(\varepsilon^2+T^2-\frac{2}{D-3}\varepsilon
T\right)}\,.
\eeqa
We may regard these as the $D$-dimensional versions of the Levi-Civita
strings. The two parameters $\varepsilon$, $T$ are the energy
density and tension measured at asymptotic infinity. They coincide with
the energy density and tension of the sources for the linearized
approximation to the solutions. When $T=\varepsilon/(D-3)$ we recover the
black string\footnote{Observe that this contains the conical-defect
strings for $D=4$.}, but in all other cases the solutions present naked
singularities where $f=0$, including in particular the strings with
Lorentz-invariant worldsheet, $\varepsilon=T$ 
\cite{Gregory:1995qh}\footnote{The case $\varepsilon=T/(D-3)$ has
conical singularities when
$-\infty<z<\infty$. Note also that only when $\varepsilon>(D-3)T$ or
$\varepsilon< T/(D-3)$ does the angular $S^{D-3}$ shrink to zero at the
singularity.}. One might nevertheless expect that, like in four
dimensions, all of these strings with $T>0$ should be able to accelerate
a massive object. The black string might be more appealing physically,
but it is not known whether it can pierce a black hole horizon in a
non-singular manner\footnote{The
methods of \cite{Emparan:2009at} should be of help here.}. 

This suggests that in $D>4$, as in four dimensions, a family of
accelerating black hole solutions should exist with at least three
independent parameters, for the black hole mass and the string energy
density and tension, with an open set of their values being potentially
useful for physical applications. But, unlike in four dimensions, the
asymptotic behavior does not seem to single out any specific solution,
so in this respect they all appear to be on a similar footing and
different pulling strings may be relevant to different problems. It is
even possible that strings with non-uniform density need to be
considered, \eg in order to satisfy the junction conditions on the brane
as suggested by the results of \cite{Kodama:2008wf}. One may also
consider struts pushing on the black holes, but they are always nakedly
singular since there are no `black struts'. 

\section*{Acknowledgments}

Work supported by DURSI 2009 SGR 168, MEC FPA 2007-66665-C02 and CPAN
CSD2007-00042 Consolider-Ingenio 2010. JC was also supported in part by
FPU grant AP2005-3120.

\appendix

\section{The solutions in $(x,y)$ coordinates}

The C-metric is customarily written not in Weyl coordinates but in a set
of coordinates $(x,y)$ adapted to uniformly accelerated motion.
In order to write our solutions in these coordinates, we perform the
change
\beqa
\rho&=&\frac{2}{A^\alpha(x-y)^2}\sqrt{(1-x^2)(y^2-1)(1+\nu x)(1+\nu
y)}\,,\nonumber\\
z&=&\frac{(1-x y)(2+\nu(x+y))}{A^\alpha(x-y)^2}\,.
\eeqa
and
\begin{equation}
a_1=-\frac{\nu}{A^\alpha}\,,\qquad a_2=\frac{\nu}{A^\alpha}\,,
\qquad a_3=\frac{1}{A^{\alpha}}\,,
\end{equation}
which can be generically applied to any Weyl solution with two finite rods.
The parameter $A$ fixes the
overall scale, and its exponent is
\beqa
\alpha&=&\frac{2}{1+m}\quad \mathrm{for~solutions~with~strings}\,,\nonumber\\
\alpha&=&2\quad \mathrm{for~solutions~with~struts}\,.
\eeqa
We take $y\leq -1$, $-1\leq x \leq 1$ and $0<\nu<1$.

The  functions $\mu_i$ in \eqref{mui} then become
\beqa
\mu_1&=&2\frac{(x-1)(1+y)(1+\nu y)}{A^\alpha(x-y)^2}\,,\nonumber\\
\mu_2&=&2\frac{(x-1)(1+y)(1+\nu x)}{A^\alpha(x-y)^2}\,,\\
\mu_3&=&2\frac{(y^2-1)(1+\nu x)}{A^\alpha(x-y)^2}\,.\nonumber
\eeqa
Defining $G(\xi)=(1-\xi^2)(1+\nu\xi)$, the metric with strings of
sec.~\ref{strings} reads
\beqa
ds^2&=&\frac{2}{A^2(x-y)^2}\left[G(y)\left(\frac{2G(x)}{(x-y)^2}\frac{1-y}{1-
x}\right)^m dt^2 +G(x)\left(\frac{2G(x)}{(x-y)^2}\frac{1-y}{1-
x}\right)^{-m} \frac{d\phi^2}{\bar{C}^2}\right]\nonumber\\
&&+
\frac{\Upsilon}{A^2}\left(-\frac{dy^2}{G(y)}+\frac{dx^2}{G(x)}\right)\,,
\eeqa
where
\begin{equation}\label{xystring}
\Upsilon=2\left(\frac{1-\nu}{x-y}\right)^2
\left[\frac{(1-x)(1-y)(2+\nu(1+x+y-xy))^{2-m}}{2G(x)} 
\left(\frac{G(x)(1-y)}{(1-x)(x-y)}\right)^m\right]^m\,,
\end{equation}
and
\begin{equation}
\bar{C}=\frac{1}{2^m}\frac{(1+\nu)^{1-m}}{1-\nu}\,.
\end{equation}

For the solution with struts of
sec.~\ref{struts} we get
\beq\label{xystrut}
ds^2=\frac{2}{A^2(x-y)^2}\left[G(y)\left(\frac{1-x}{1-y}\right)^m 
dt^2+G(x)\left(\frac{1-x}{1-y}\right)^{-m} d\phi^2\right]+
\frac{\Upsilon}{A^2}\left(-\frac{dy^2}{G(y)}+\frac{dx^2}{G(x)}\right)\,,
\eeq
now with
\beq
\Upsilon=2\left(\frac{1-\nu}{x-y}\right)^2\left(
\frac{\left[-(x+y+\nu(1+x y))(2+\nu(-1+x+y+x y))\right]^{2-m}}
{4(1-\nu)^{2(2-m)}\left((1-x)(1-y)\right)^{1-m}}\right)^m\,.
\eeq

When $m=0$ both solutions reduce,
up to a constant rescaling of coordinates, to the uncharged C-metric with
the factorized form for $G(\xi)$ first given in \cite{Emparan:1996ty}
(see also \cite{Hong:2003gx}).

\end{document}